\documentclass[a4paper,10pt,twoside]{cpc-hepnp}
\usepackage{CJK,upgreek,fancyhdr}
\usepackage{multicol}
\usepackage{graphicx}
\usepackage{booktabs}
\usepackage{amssymb,bm,mathrsfs,bbm,amscd}
\usepackage[tbtags]{amsmath}
\usepackage{lastpage}

\begin{document}
\begin{CJK*}{GB}{gbsn}

\fancyhead[c]{\small Chinese Physics C~~~Vol. xx, No. x (201x) xxxxxx}
\fancyfoot[C]{\small 010201-\thepage}

\footnotetext[0]{Received xx xx 201x}

\title{Plastic scintillation detectors for precision time-of-flight measurements of relativistic heavy ions \thanks{Supported by National Natural Science Foundation of China (11105010,11235002,11475014), and National Program on Key Basic Research Project (2016YFA0400502)}}

\author{%
 Wen-Jian Lin$^{1}$%
\quad Jian-Wei Zhao$^{1}$%
\quad Bao-Hua Sun$^{1;1)}$\email{bhsun@buaa.edu.cn}%
\quad Liu-Chun He$^{1}$%
\quad Wei-Ping Lin$^{2}$%
\quad Chuan-Ye Liu$^{1}$%
\quad Isao Tanihata$^{1}$
\\
\quad Satoru Terashima$^{1}$%
\quad Yi Tian$^{1}$%
\quad Feng Wang$^{1}$%
\quad Meng Wang$^{1}$%
\quad Guang-Xin Zhang$^{1}$%
\quad Xue-Heng Zhang$^{2}$%
\\
\quad Li-Hua Zhu$^{1}$%
\quad Li-Min Duan$^{2}$%
\quad Rong-Jiang Hu$^{2}$%
\quad Zhong Liu$^{2}$%
\quad Chen-Gui Lu$^{2}$%
\quad Pei-Pei Ren$^{2}$%
\\
\quad Li-Na Sheng$^{2}$%
\quad Zhi-Yu Sun$^{2}$%
\quad Shi-Tao Wang$^{2}$%
\quad Tao-Feng Wang$^{1}$%
\quad Zhi-Guo Xu$^{2}$%
\quad Yong Zheng$^{2}$%
}
\maketitle

\address{%
$^1$ School of Physics and Nuclear Energy Engineering, Beihang University, Beijing 100191, China\\
$^2$ Institute of Modern Physics, Chinese Academy of Sciences, Lanzhou 730000, China\\
}

\begin{abstract}
Plastic scintillation detectors for Time-of-Flight (TOF) measurements are almost essential
for event-by-event identification of relativistic rare isotopes. In this work, a pair of plastic scintillation detectors of 50 $\times$ 50 $\times$ 3$^{t}$ mm$^3$ 
and 80 $\times$ 100 $\times$ 3$^{t}$ mm$^3$ have been set up at the external target facility (ETF), Institute of Modern Physics. 
Their time, energy and position responses are measured with $^{18}$O primary beam at 400 MeV/nucleon. After the off-line walk-effect and position corrections, the time resolution of the two detectors are determined to be 27 ps ($\sigma$) and 36 ps ($\sigma$), respectively. Both detectors have nearly the same energy resolution of 3$\%$ ($\sigma$) and position resolution of 2 mm ($\sigma$). The detectors have been used successfully in nuclear reaction cross section measurements, and will be be employed for upgrading RIBLL2 beam line at IMP as well as for the high energy branch at HIAF.
\end{abstract}

\begin{keyword}
Plastic scintillator, Time-of-Flight, time resolution, walk-effect correction
\end{keyword}

\begin{pacs}
29.27.Ac, 29.40.Mc, 29.85.+c
\end{pacs}

\footnotetext[0]{\hspace*{-3mm}\raisebox{0.3ex}{$\scriptstyle\copyright$}2013
Chinese Physical Society and the Institute of High Energy Physics
of the Chinese Academy of Sciences and the Institute
of Modern Physics of the Chinese Academy of Sciences and IOP Publishing Ltd}%

\begin{multicols}{2}

\section{Introduction}

Plastic scintillation detectors with fast timing are almost essential for Time-of-Flight (TOF) determinations of charged particles at relativistic energies of a few hundreds of MeV/nucleon, aiming for unambiguously particle identification (see e.g.~\cite{Voss1995,J.PLB}). Taking the nuclear fragmentation as an example, the reaction products in this process retain to keep their initial velocities as the incident ions. This means that on a few meters distance the TOF differences between reaction products is only on the order of a few ns.

To improve the TOF resolution, a long flight path length of several tens of meters is generally needed to separated different ions. 
An alternative way to enhance the TOF resolution is clearly to improve the precision in time determination of TOF. This sometimes gets more important due to the limited flight path. The improved TOF resolution also open many new opportunities, and one of which is the direct mass measurement of exotic nuclei~\cite{Lunney03,Matos-NIMA2012,Meisel2013,Sun:2015vza,Sun:2016hmk}.

Recent development of plastic scintillators with fast decay times and high-speed photomultiplier tubes (PMT) provide us the opportunity to perform fast timing measurements 
with a resolution down to about 10 ps ($\sigma$). To achieve such goal, the size of plastic scintillator has to minimize, due to the fact that the time resolution significantly degrades when increasing the scintillator's length. For instance, 15$\times$25.4$\times$0.254$^{\rm t}$ mm$^3$~\cite{Matos-NIMA2012}, 30$\times$20$\times$0.5$^{\rm t}$ mm$^3$~\cite{Nishimura03}, and 30$\times$10$\times$3$^{\rm t}$ mm$^3$~\cite{Zhao:2016pja} have been used to reach 10 ps resolution, respectively.  
In reality, such size detectors are installed at the focal planes of fragment separators where the beam is focused and centered well to a small size. 

In current radioactive beam experiments, the secondary heavy ions have typically a relatively large spread in both space and angle. Meanwhile, in many reaction studies, a parallel beam 
are preferred. In these cases, a plastic scintillator with size of 50$\times$50 mm$^2$ or larger are required to fully cover the beam size of secondary ions for particle counting and identification. 
Standard technique of utilizing such size plastic scintillators read out by two or four photomultiplier tubes, can give a typical time resolution of about 100+ ps, which is sufficient to discriminate light relativistic radioactive ion species~\cite{Leo}. However, when going to heavier system with $A>100$ of interest, 
a time resolution of less than 50 ps is required in particular for short flight path, such as RIBLL2 at IMP. 

As a natural extension to the conventional scheme with two or four high-precision measurements of the same physical event, it is possible to gain better time resolution by increase the active area covered by PMT area. This approach has been tested recently using a circular (27 cm in diameter) BC-420 plastic-scintillator sheet read-out by 32 photomultiplier tubes and a time resolution 
down to the order to 10 ps (1 $\sigma$) has been achieved~\cite{Ebran201340}. 

For design of large size plastic scintillation detectors, the thickness of plastic scintillator is one of the critical parameters. 
Although less material in the beam line is generally required to avoid extra interactions and energy straggling of the incident beam, a worse time resolution is expected with a thin scintillator 
due to less number of photon emission and thus smaller pulse height. Indeed, as the thickness of scintillator increased the time resolution is improved accordingly~\cite{Nakajima20084621,Ebran201340}.   
Moreover, a combination of time-to-analog converter (TAC) with anolog-to-digital converter (ADC) currently has an advantage over time-to-digital converter (TDC) for signal digitization due to
a better sampling resolution. Attempts~\cite{Li:2013zza, Wu:2008} were also made by replacing the conventional electronics by waveform digitizers, which have the clear advantages in recording the full event signals but so far are still limited by their sampling rates.

In this work, we report a new investigation of plastic scintillation detectors with relatively large sizes. They satisfy the requirement for detecting relativistic heavy ions: 
large spreads in space, angle and energy deposited, and excellent timing for particle identification of heavy ions, 
while meanwhile minimizing the number of PMTs and electronics used.  
This work is complementary to previous investigations~\cite{Nishimura03,Matos-NIMA2012,Nakajima20084621,Hoischen2011,Ebran201340,Zhao:2016pja} in thicknesses and sizes of scintillators, types and energies of primary beams, and total energy losses in plastic scintillators, but using the same types of plastic scintillators and PMTs. Especially, 
we present the main procedure and details on how to improve the time resolution using the valuable pulse-height and position information. 
The detectors have been already successfully employed for nuclear charge-changing cross section measurements of exotic nuclei, in which a large coverage of reaction products are crucially important.
Similar detectors will be used for upgrading the RIBLL2 beam line at IMP, and be the key component for the future high-energy fragment separator at HIAF. 

This paper is organized as follows. The experiment is introduced in Section 2. The results and the data analyses on the TOF resolution, energy resolution and position resolution are presented in detail in Section 3.  Finally, a summary is given in Section 4.

\section{Experiment}

The experiment was performed at the Heavy Ion Research Facility in Lanzhou (HIRFL)~\cite{CSR-Xia200211}. The primary $^{18}$O beam at 400 MeV/nucleon was extracted from the the synchrotron cooler storage main ring (CSRm), and then was separated according to magnetic rigidity by using the Second Radioactive Ion Beam Line in Lanzhou (RIBLL2).

The layout of the apparatus is shown in Fig. \ref{Fig.1}. The detectors was mounted at External Target Facility (ETF)~\cite{ETF2014-PhysRevC.90.037601}. One plastic scintillator (PL0) with a cross-section of 100 $\times$ 100 mm$^2$ and a thickness of 3 mm installed at F1 was used as the TOF start. This start detector was designed based on EJ200 plastic scintillator bar. One photomultiplier tube (PMT)
of Hamamatsu R7111 was used to give the signals from one end of the scintillator.

\begin{center}
\includegraphics[width=8cm]{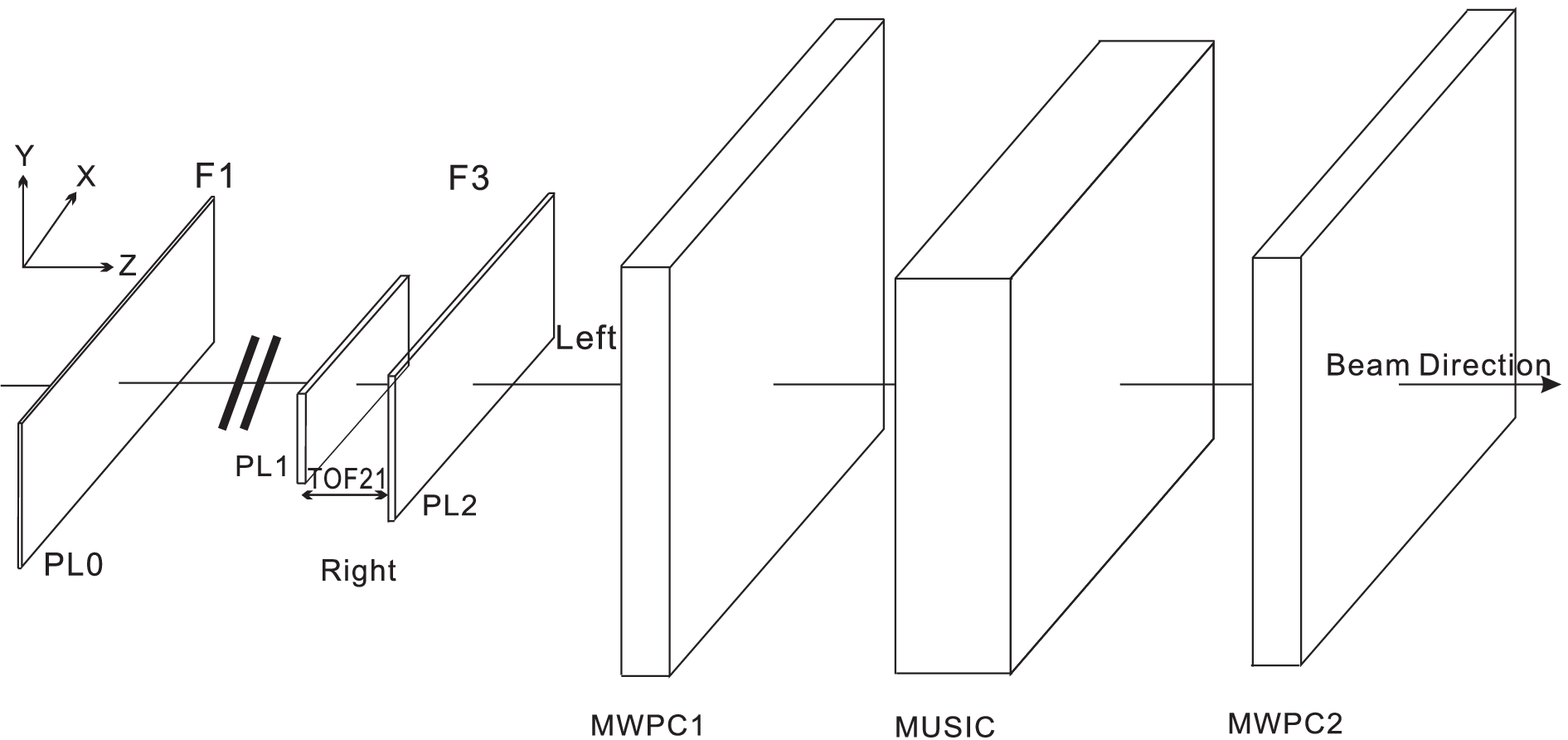}
\figcaption{\label{Fig.1}   Schematic view of the experimental setup for the TOF test experiment at RIBLL2. }
\end{center}

The plastic scintillation detectors (PL1 and PL2) of test in the present work were installed at F3, about 26 meters away from PL0~\cite{Xue-Heng:2013ica}. Both ends of PL1 and PL2 were coupled with PMTs of type H2431 through optical silicone rubber EJ-560. As for the corresponding plastic scintillators, PL1 is EJ-232Q while PL2 is EJ-232. EJ-232Q (with 0.5\% quenching level of benzophenone) is a quenched version of EJ-232. The sizes of PL1 and PL2 are 50 $\times$ 50 mm$^{2}$ and 80 $\times$ 100 mm$^2$, respectively. The thicknesses of them are all fixed to 3 mm. PL1 was also used as the trigger detector for this work. The scintillators were wrapped with aluminum foils, and then with black plastic fabric for light-tight.

The position information of each ion was determined by two multi-ware proportional chambers (MWPC1 and MWPC2) placed after the plastic scintillation detectors, while in between a multiple sampling ionization chamber (MUSIC)~\cite{Zhang2015389} was installed to get the energy deposited of particles of interest. All the $x$ position information shown below is deduced from MWPC1 and MWPC2.

Signals from each PMT were split into two, to provide both energy and time information. One was delivered to a Charge-to-Digital Converter (QDC) CAEN v792 for energy loss measurement. The other signal was first fed into a leading edge discriminator (LED) CAEN v895, then again split into two: one to the Time-to-Digital Converter (TDC) CAEN v775 and the other to the  Multi-Hit/Multi-Event Time to Digital Converter (MULTITDC) CAEN v1290. The discriminator thresholds of CAEN v895 were set as low as possible but above the noise level of PMTs. To account for the long TOF distance from F1 to F3, the full range of TDC is set to 280 ns, corresponding to 68.5 ps per channel. The time range calibration of TDC was done channel by channel using ORTEC 462 and GG8020. The MULTITDC was used for cross check and its full scale range is set to 4 $\mu s$, corresponding to 25 ps per channel.

\section{Data analyses and Results}

The timing is obtained by taking the average from left PMT and right PMT (see Fig. \ref{Fig.1}). For example, the raw timing for PL1 ($T_{1raw}$) is eventually calculated as

\begin{equation}
   T_{1raw} = (T_{1L}+T_{1R})/2, \;
\end{equation}
where $T_{1L}$ and $T_{1R}$ are the timing determined from the left and right side PMT, respectively.

For each ion hitting on scintillator bars, the integrated charges $Q_L$ ($Q_R$) obtained from the left (right) side of plastic scintillators, 
is proportional to the relational expression of the total energy loss $Q_0$ and the hitting position $x$:
\begin{equation}
\begin{aligned}
 &Q_L\propto Q_0e^{-\lambda_{L} x},\\
 &Q_R\propto Q_0e^{-\lambda_{R} (L-x)}.\\
\end{aligned}
\end{equation}
Here $\lambda_{L}$ and $\lambda_{R}$ represent the light transmission parameters that depend on the left and right parts of the detector, respectively. $L$ is the full length of the plastic scintillation bar. In this work, it is found that $\lambda_L \approx \lambda_R$. In such case, the deposited energy information in the plastic scintillator is then evaluated as the geometrical mean value $\bar{Q}$:
\begin{equation}
  \bar{Q} = \sqrt{Q_L\cdot Q_R}. \;
  \label{eq:energy}
\end{equation}
Care should be taken to evaluate the energy deposited in the scintillator if $\lambda_{L}$ and $\lambda_{R}$ are significantly different, because $\bar{Q}$
here is position dependent. 

\subsection{Time resolution}

\subsubsection{Time-walk effect and position correction}

For precise determinations of the intrinsic time resolution of plastic scintillation detectors, one needs a careful investigation of the effect coming from the time walk due to the variation of pulse amplitudes, and the other inducement due to different hitting positions on the plastic scintillators. Both effects can have substantial influence on the time resolution. Previous investigations found that simultaneously measurements of both time and pulse-height are very valuable for correction of time-walk effect~\cite{Nishimura03,Zhao:2016pja}. 

Taking PL2 as an example, the time dependence on the pulse height can be clearly seen in Fig. \ref{Fig.2}. To correct this time-walk effect, we use the following formula proposed in Ref.~\cite{Zhao:2016pja}:
\begin{equation}
T=T_{raw}-\frac{c_{raw}}{\sqrt[4]{Q_L\cdot Q_R}},
\end{equation}
where $T_{raw}$ and $T$ are the measured timing and the corrected timing with pulse-height, respectively. $C_{raw}$ is the walk-effect correction coefficient. It was determined by using only those events hitting onto a fixed $x$ position of plastic scintillator. The plot after walk-effect correction is shown in Fig.~\ref{Fig.3}. Clearly, a large portion of pulse-height dependence can be removed.

\begin{center}
\includegraphics[width=8cm]{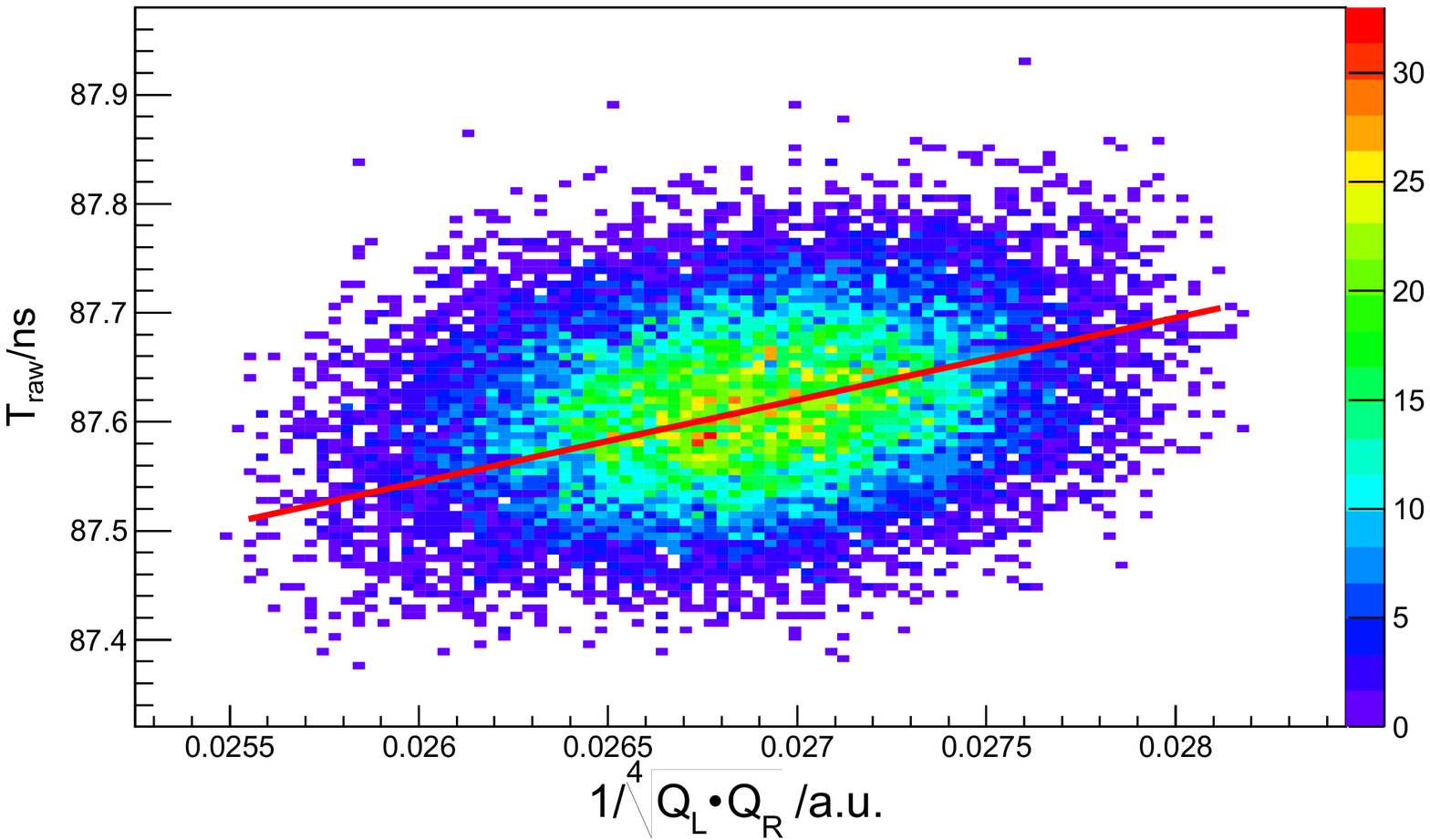}
\figcaption{\label{Fig.2}   Pulse-height dependence of the timing. The red line indicates the linear fitting. }
\end{center}

\begin{center}
\includegraphics[width=8cm]{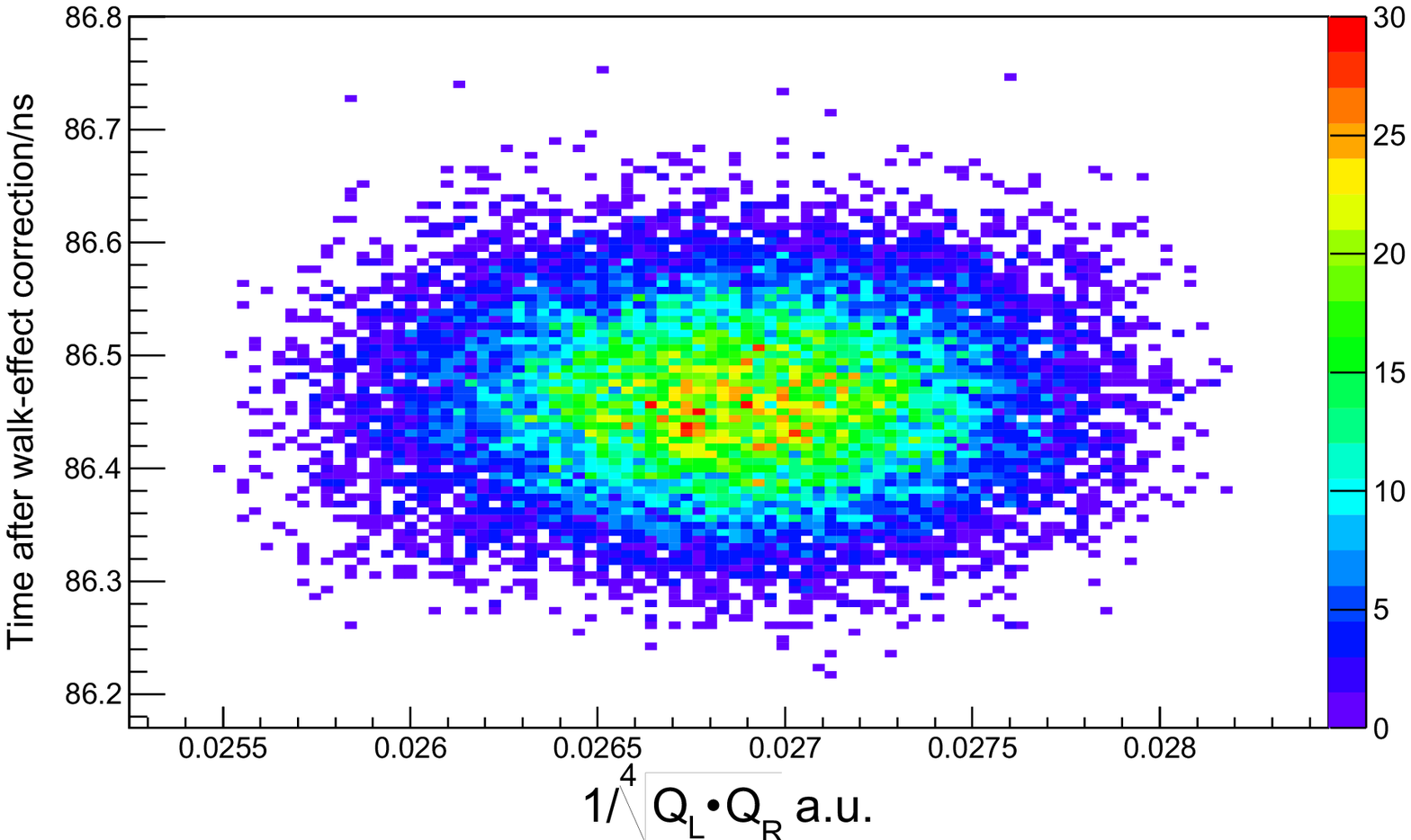}
\figcaption{\label{Fig.3}   Same as Fig. \ref{Fig.2} but after walk-effect correction. }
\end{center}

However, as displaced in Fig. \ref{Fig.4}, a distinct position dependence is still seen after walk-effect correction. Considering the fact that ions hitting on different positions will result in a different travel time towards both PMTs, the following equation has been applied to eliminate such effect~\cite{Wu:2005xk}:

\begin{equation}
T^{'}=T-a\cdot x,
\end{equation}
where $T^{'}$ is the time after position correction and $a$ is the correction parameter to be determined from the experimental data.

\begin{center}
\includegraphics[width=8cm]{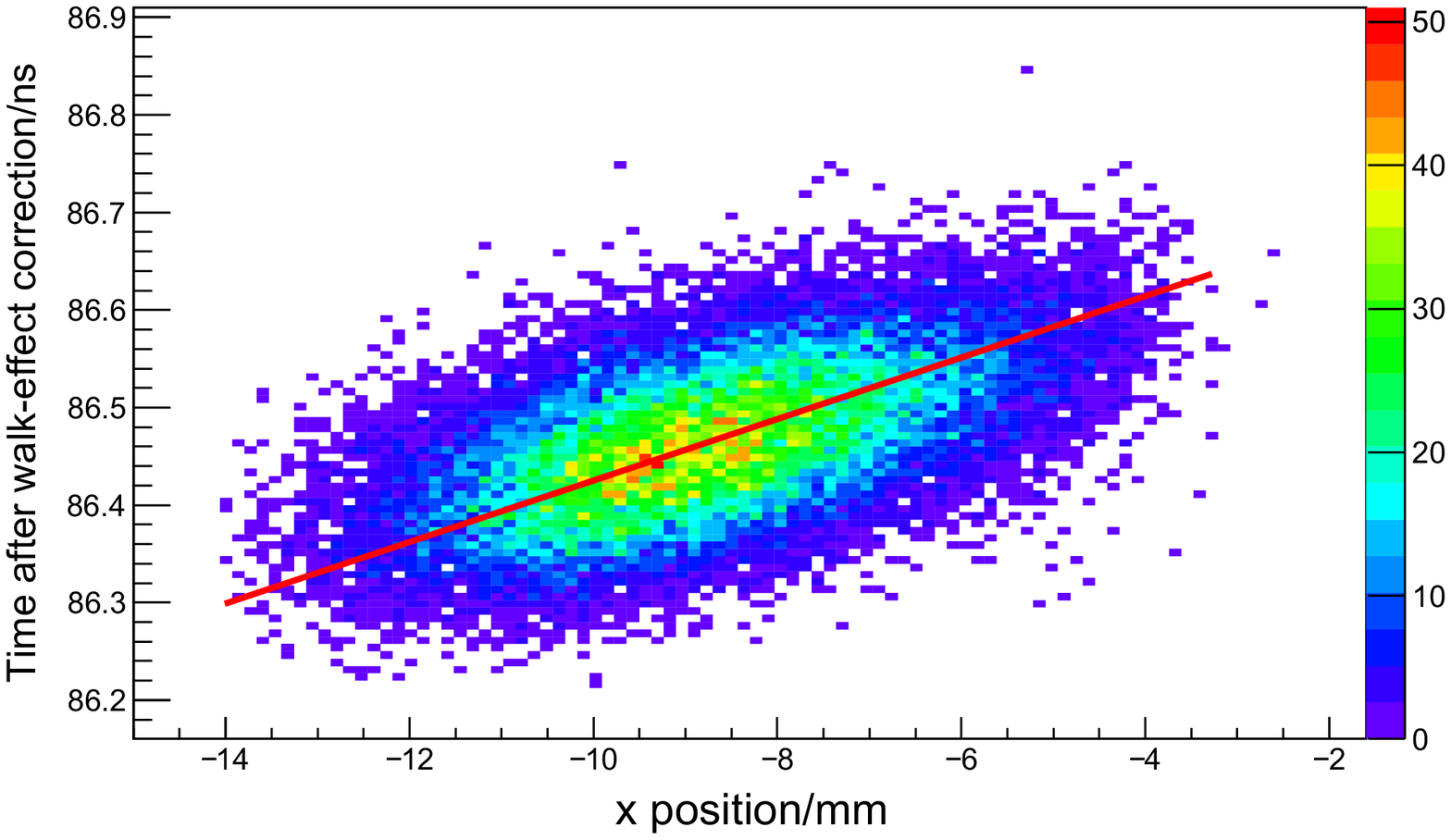}
\figcaption{\label{Fig.4}   Position dependence of the timing. The red line indicates the linear fitting. }
\end{center}

With both walk-effect and position correction, Fig. \ref{Fig.5} shows that the timing now has no dependence on neither the variation of pulse amplitude nor the hitting positions. From this we can extract the intrinsic time resolution of the detector, including the contribution of the plastic scintillator, PMT and electronics.

\begin{center}
\includegraphics[width=8cm]{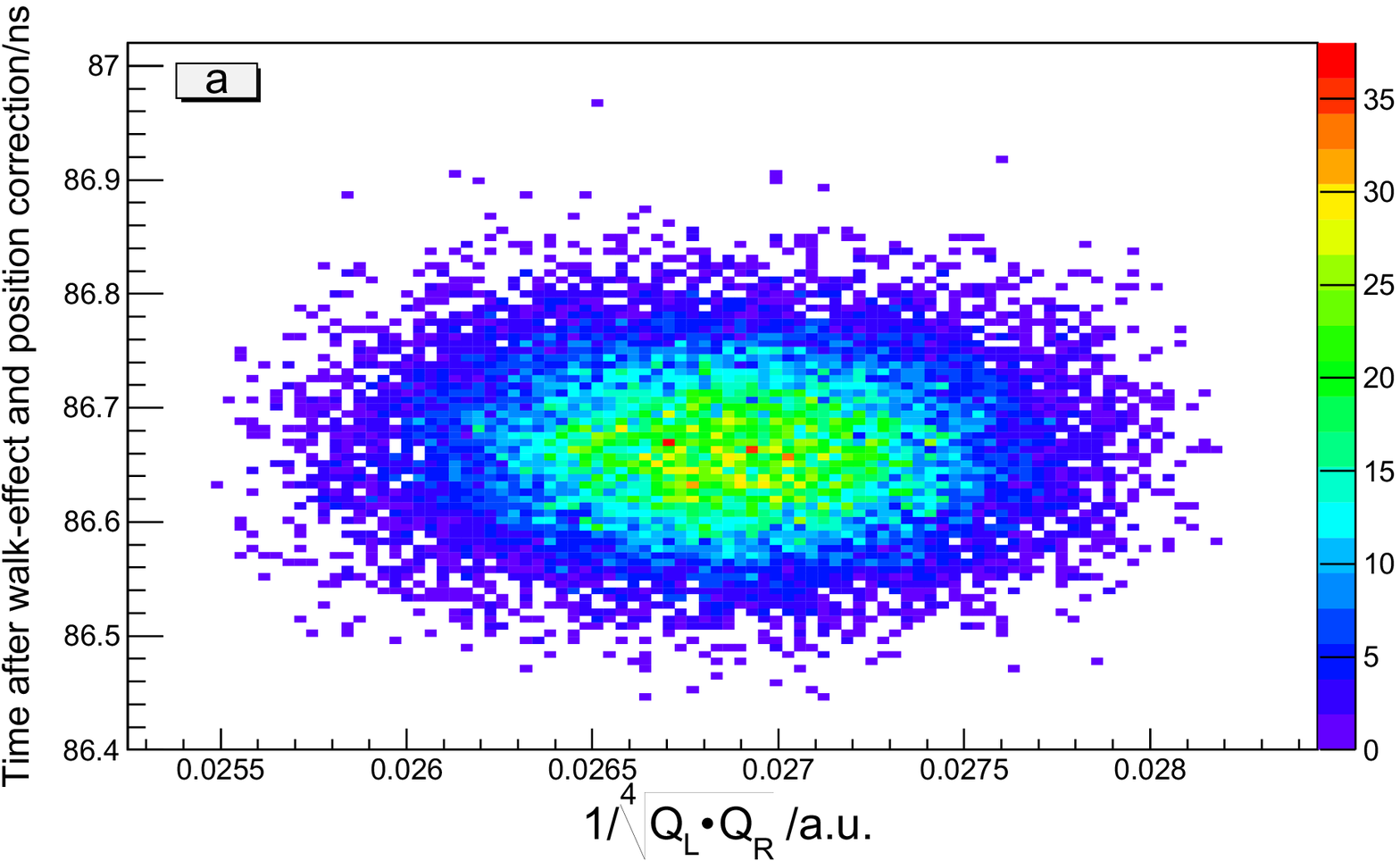}
\end{center}

\begin{center}
\includegraphics[width=8cm]{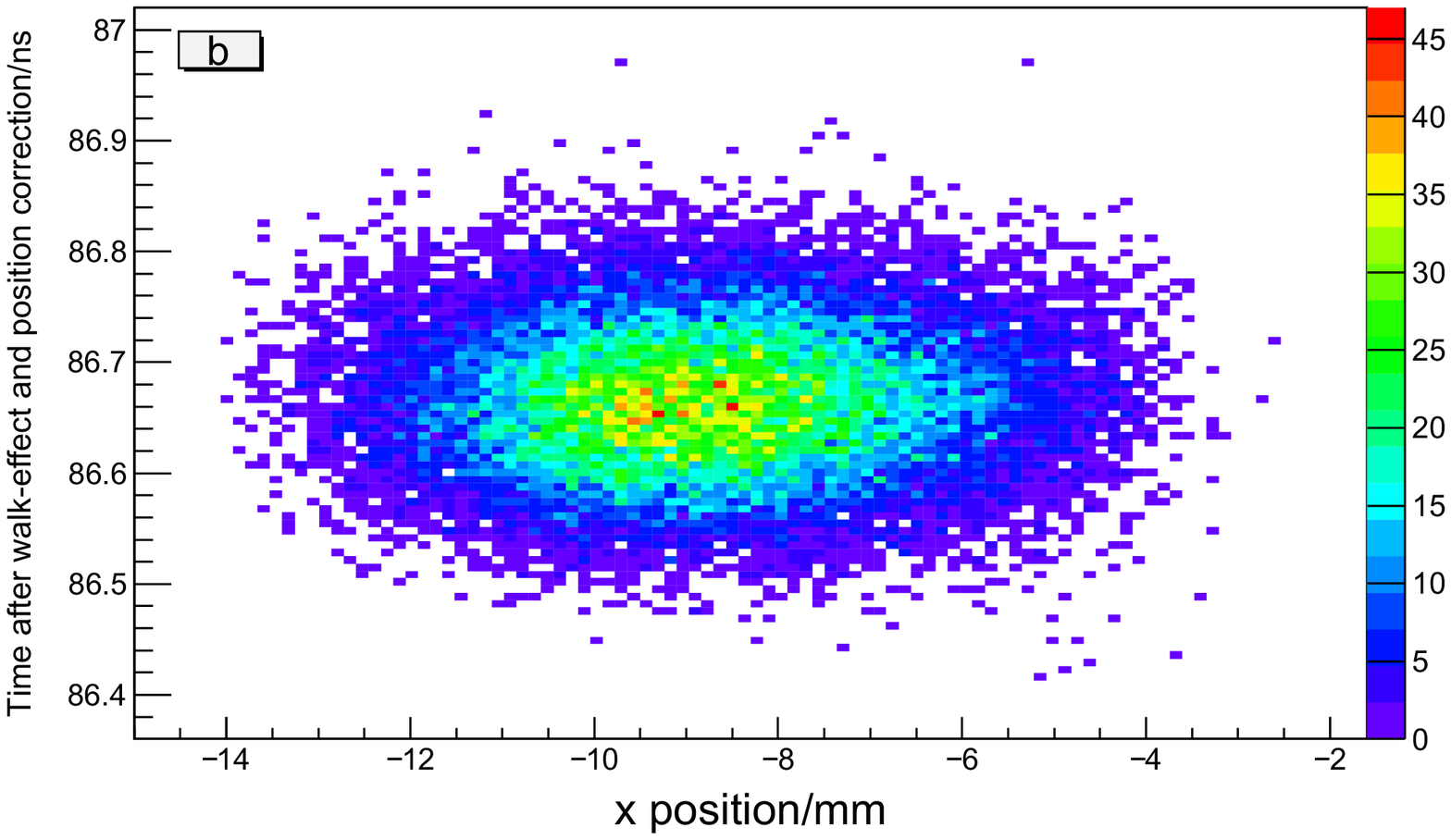}
\figcaption{\label{Fig.5} Timing after walk-effect and position correction as a function of pulse-height (a) and hitting position (b). }
\end{center}

\subsubsection{TOF measurements}

TOF distribution can be described very well by a Gaussian function. Taking TOF21 defined as $T_{2}-T_{1}$ as an example, Fig. \ref{Fig.6} shows the Gaussian fits of TOF21 distributions in cases without correction, with only walk-effect correction and with both walk-effect and position correction.

\begin{center}
\includegraphics[width=8cm]{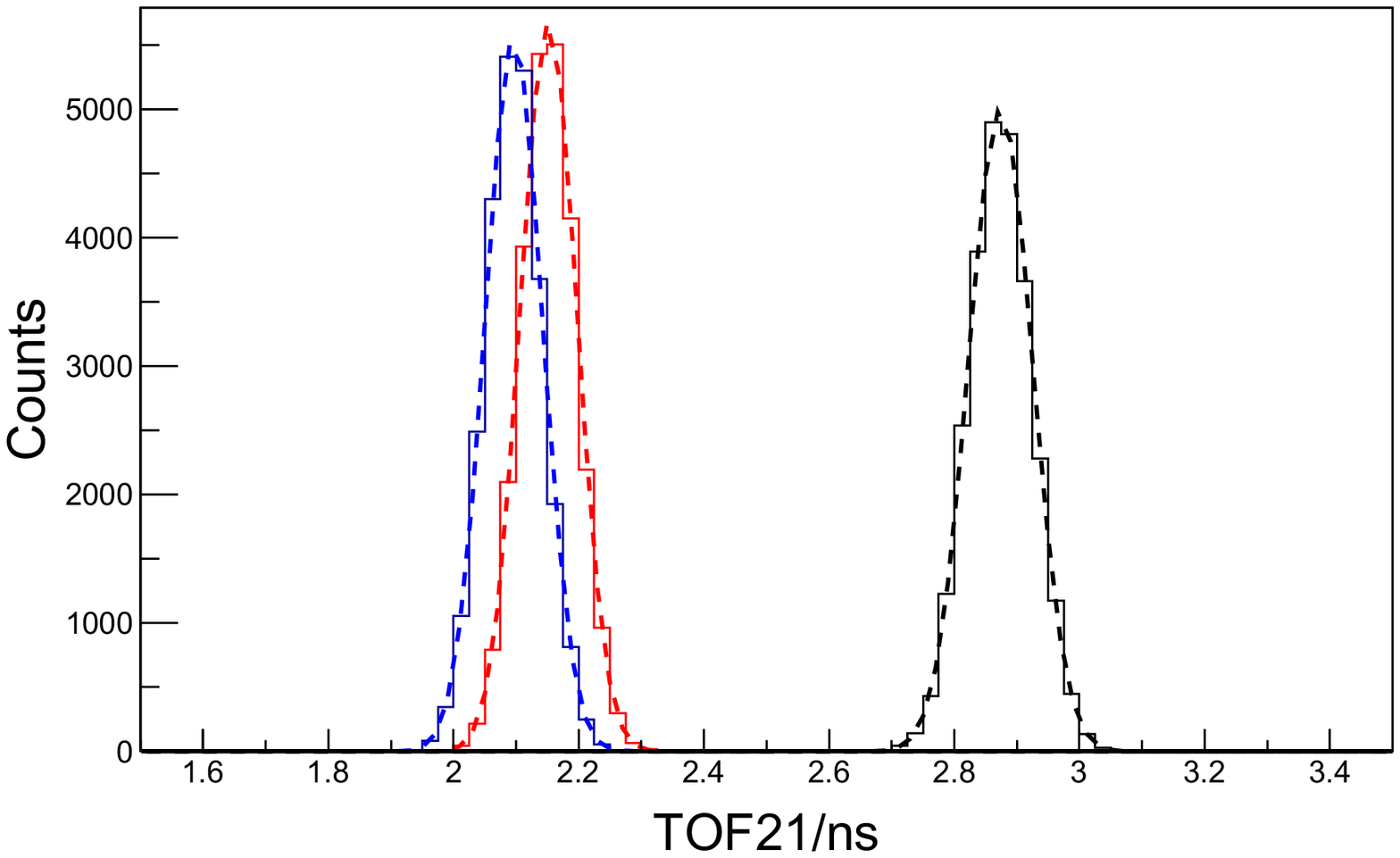}
\figcaption{\label{Fig.6}   TOF21 distributions in cases without correction (black line), with only the walk-effect correction (blue line) and with both walk-effect and position correction (red line). The Gaussian fits (dashed lines) for the three cases show a standard deviation ($\sigma$) of 51 ps, 46 ps and 45 ps, respectively.}
\end{center}

To determine the time resolution of PL1 and PL2, TOF21, TOF10 and TOF20 are needed. For the one side readout detector like PL0, we use the following walk-effect correction formula~\cite{Nishimura03, BRAUNSCHWEIG1976261, TANIMORI198357}:
\begin{equation}
T=T_{raw}-\frac{C}{\sqrt{Q}},
\end{equation}
where $C$ is the correction parameter to be determined.
After both correction, the resolution of TOF21, TOF10 and TOF20 are obtained as 45 ps ($\sigma$), 83 ps ($\sigma$) and 87 ps ($\sigma$), respectively.
It is found that the position correction has only little influence on TOF10 and TOF20. This is limited by the relatively poor time resolution of PL0.

The time resolution of PL1 and PL2 can be calculated by the following formula:
\begin{equation}
\begin{aligned}
\sigma_{12}^{2}=\sigma_{1}^{2}+\sigma_{2}^{2},\\
\sigma_{10}^{2}=\sigma_{1}^{2}+\sigma_{0}^{2},\\
\sigma_{20}^{2}=\sigma_{2}^{2}+\sigma_{0}^{2},\\
\end{aligned}
\end{equation}
where $\sigma_{12}$, $\sigma_{10}$ and $\sigma_{20}$ stand for the Gaussian fitting standard deviation $\sigma$ of TOF21, TOF10 and TOF20. $\sigma_{1}$, $\sigma_{2}$ and $\sigma_{0}$ represent the time resolution of PL1, PL2 and PL0, respectively. The determined resolution for three detectors are 27 ps ($\sigma$), 36 ps ($\sigma$) and 79 ps ($\sigma$), respectively.

To verify this result, the data registered by MULTITDC are analyzed as well. As shown in Fig.~\ref{Fig.7},  50 ps resolution ($\sigma$) is obtained by using the MULTITDC data. This is fully consistent with that from TDC (see Fig. 6).

\begin{center}
\includegraphics[width=8cm]{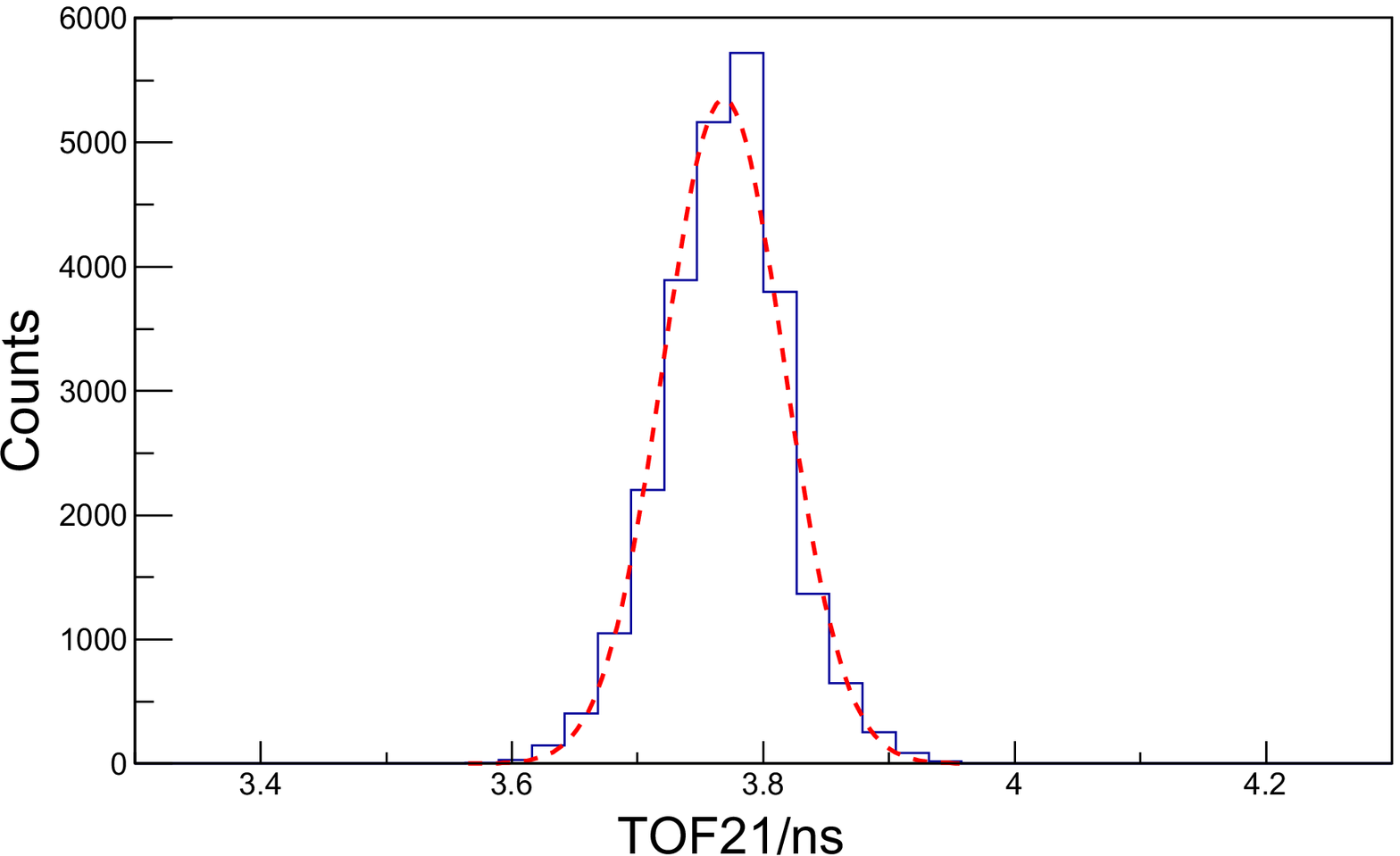}
\figcaption{\label{Fig.7} TOF21 distribution (solid line) using the data registered by MULTITDC. A Gaussian fit is indicated by the dashed line.}
\end{center}

\subsection{Energy resolution}

The energy loss of each ion in the detector can be described by Eq. 3. Shown in Fig. \ref{Fig.8} is the energy resolution of PL1.

\begin{center}
\includegraphics[width=8cm]{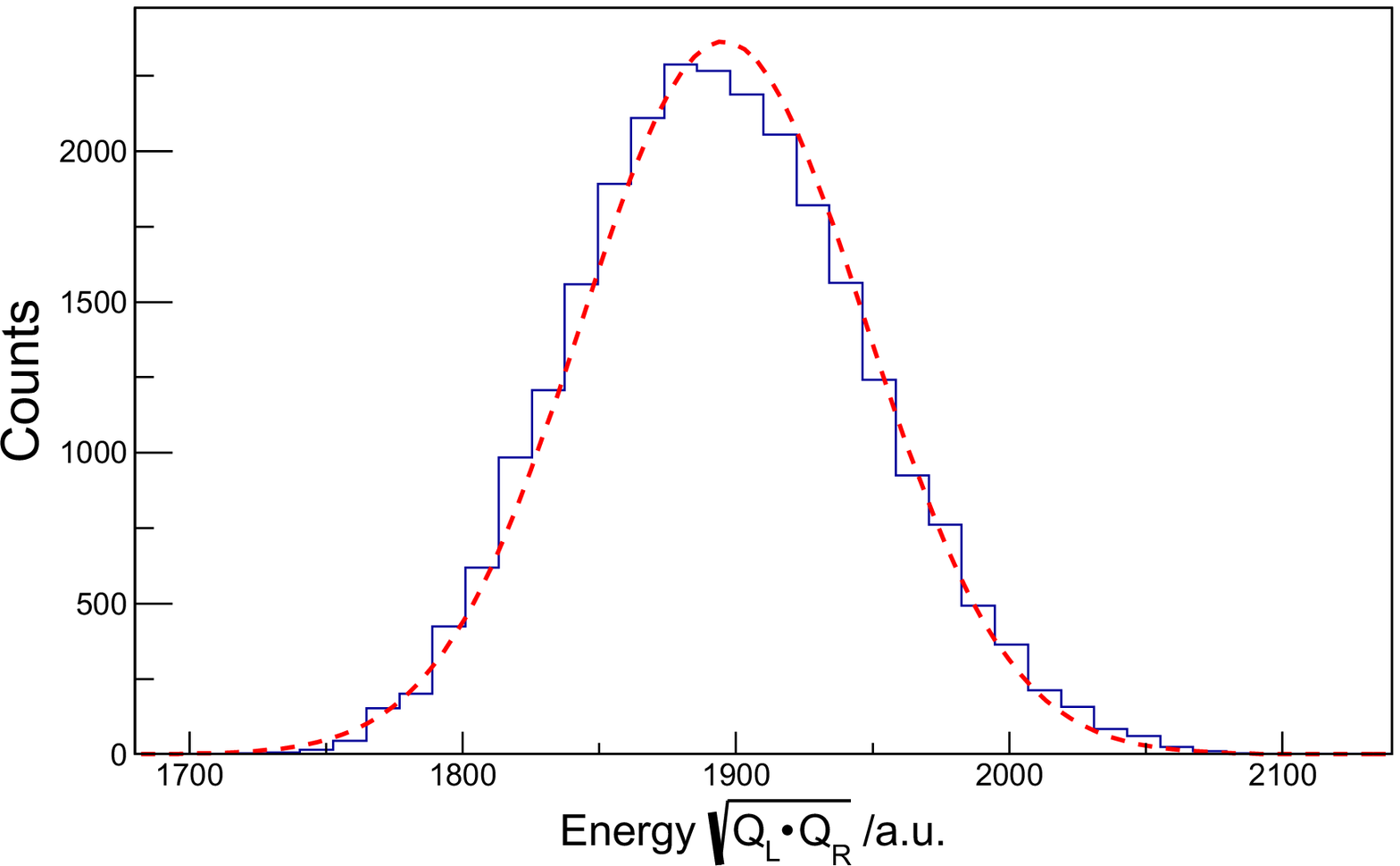}
\figcaption{\label{Fig.8}   Energy resolution of PL1 is about $3\%$ ($\sigma$) deduced by Gaussian fit. }
\end{center}

It is found that both PL1 and PL2 have nearly the same energy resolution of $3\%$ ($\sigma$). This is almost comparable to a large size Si detector and ionization chamber. Such a fast plastic scintillator can be useful for both timing and energy determination identification for light ions.

\subsection{Position resolution}

The position ($x$) is usually obtained from the time difference between $T_L$ and $T_R$,
\begin{equation}
  x \propto (T_L-T_R). \;
\end{equation}
However, as we already mentioned in Section 2, to match the long TOF pathlength and also the dynamic range of MWPC of about 160 ns, the TDC step was set to 68.5 ps. Thus position correlation with time difference is not precise enough for the position information.

On the other hand, following the Eq. 2, the horizontal position $x$ of hitting point can be obtained from the integrated charges of the left and right PMT signals, $Q_L$ and $Q_R$, by
\begin{equation}
  x  \propto \mbox{ln}(Q_L/Q_R). \;
\end{equation}

\begin{center}
\includegraphics[width=8cm]{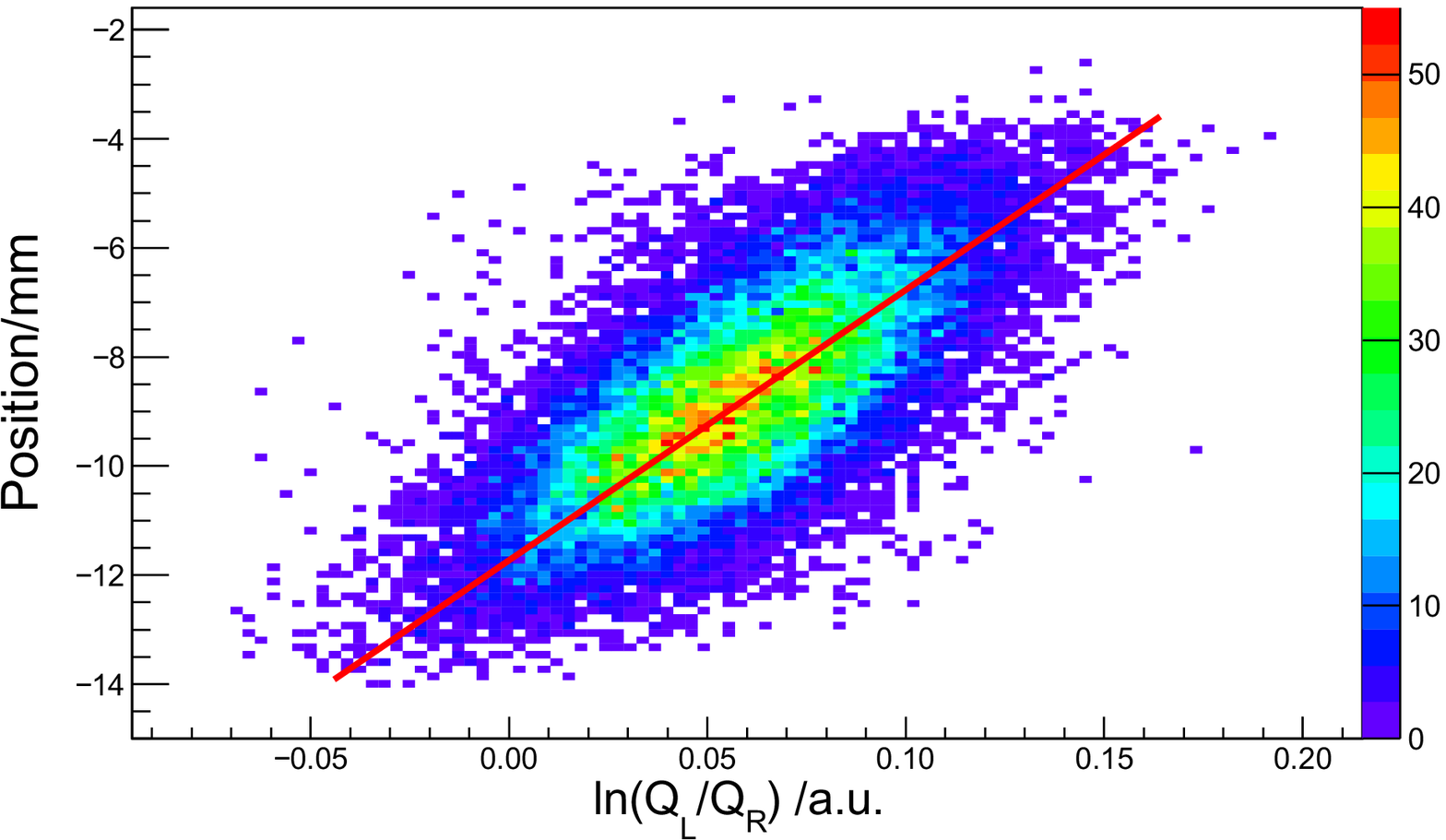}
\figcaption{\label{Fig.9}   Correlation between position and $\mbox{ln}(Q_L/Q_R)$. The linear correlation is indicated by the red line. }
\end{center}

PL1 and PL2 have nearly the same position resolution. Fig. \ref{Fig.9} shows the correlation between the $x$ position extrapolated from MWPCs and $\mbox{ln}(Q_L/Q_R)$ of PL1.
The position resolution of MWPCs is about 1 mm ($\sigma$).

Clearly, the position information is also limited by the precision of QDC. In this work, CAEN QDC v792 is used and it was set to 10 fC per channel. Plotted in Fig. \ref{Fig.10} are the corresponding $\mbox{ln}(Q_L/Q_R)$ distributions when two groups of ions with positions separated by 5 mm. At this case, they are just possible to be resolved from each other. In other words, only for ions with hitting positions separated by more than 5 mm, it is then possible to distinguish them by $\mbox{ln}(Q_L/Q_R)$. This resolution includes the contribution from both plastic scintillators and the intrinsic resolution of MWPCs. Considering the MWPC resolution of 1 mm ($\sigma$), a position resolution of 2 mm ($\sigma$) can be obtained by using the integrated charge information.

\begin{center}
\includegraphics[width=8cm]{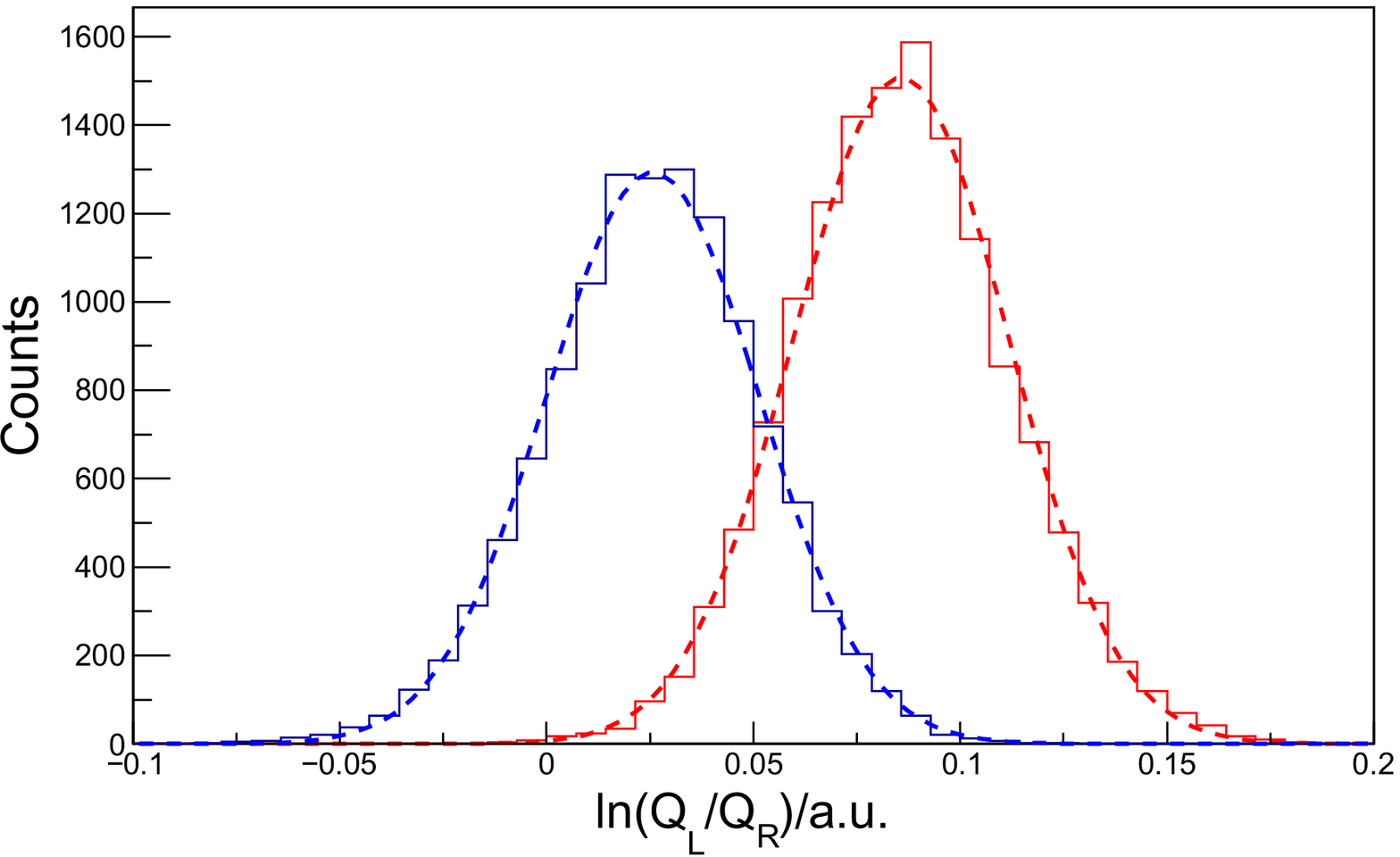}
\figcaption{\label{Fig.10}   $\mbox{ln}(Q_L/Q_R)$ distribution when $x$ positions determined by MWPCs are in the range of  -14 mm to -9 mm (blue), and -9 mm to -4 mm (red).
Gaussian fits (dashed line) of the experimental data (solid line) are shown.}
\end{center}

\section{Summary}

To summarize, a pair of plastic scintillation detectors with size of 50$\times$50$\times$3$^t$ mm$^3$ and 80$\times$100$\times$3$^t$ mm$^3$ were tested with $^{18}$O primary beam at 400 MeV/nucleon. Time resolutions of 27 ps ($\sigma$) and 36 ps ($\sigma$) are obtained after walk-effect and position correction for these two detectors, respectively. The results are verified by using the MULTITDC as a cross check. Meanwhile an energy resolution of 3$\%$ ($\sigma$) and a position resolution of 2 mm ($\sigma$) are obtained using the conventional electronics. This makes 
fast plastic scintillators possible for simultaneous measurements of energy deposited, accordingly the charge number, and position of incident ions with reasonable precisions. 
Such detectors will be used for upgrading the RIBLL2 beam line at IMP and are the key component for the high-energy fragment separator at HIAF. 

\acknowledgments{The authors would like to thank the RIBLL collaboration and the HIRFL-CSR team for providing a stable beam and assistance during the experiments.}

\end{multicols}

\vspace{-1mm}
\centerline{\rule{80mm}{0.1pt}}
\vspace{2mm}

\begin{multicols}{2}

\end{multicols}
\clearpage
\end{CJK*}
\end{document}